\documentclass{nature}

\usepackage{graphicx}
\makeatletter
\let\saved@includegraphics\includegraphics
\AtBeginDocument{\let\includegraphics\saved@includegraphics}
\renewenvironment*{figure}{\@float{figure}}{\end@float}
\makeatother

\usepackage{bm}
\usepackage{lineno}

\title{One family of 13315 stable periodic orbits of the non-hierarchical unequal-mass triple system}

\author{Xiaoming Li$^{1,2}$, Xiaochen Li$^{3,4}$ \& Shijun Liao$^{5,6}$}

\begin{document}
\maketitle

\begin{affiliations}

 \item MOE Key Laboratory of Disaster Forecast and Control in Engineering. School of Mechanics and Construction Engineering, Jinan University, Guangzhou 510632,
China
 \item Department of Earth, Atmospheric and Planetary Sciences,
Massachusetts Institute of Technology, Cambridge, Massachusetts 02139,
USA
 \item School of Civil Engineering and Transportation, South China University of Technology, Guangzhou 510641, China
 \item Department of Engineering Science, University of Oxford, Oxford OX1 3PJ, UK
 \item Center of Advanced Computing, School of Naval Architecture, Ocean and Civil Engineering, Shanghai Jiaotong University, Shanghai 200240, China 
 \item School of Physics and Astronomy,  Shanghai Jiaotong University, Shanghai 200240, China 

\end{affiliations}

\begin{abstract}
The three-body problem has been studied for more than three centuries\cite{Newton1687, Musielak2014}, and has received much more attention in recent years\cite{Jankovic2016, Archibald2018, Stone2019}. It shows complex dynamical phenomena due to the mutual gravitational interaction of the three bodies. Triple systems are common in astronomy, but all observed periodic triple systems are hierarchical up till now\cite{Reipurth2012, Dimitrov2017, Torres2019}.  
It is traditionally believed that bound non-hierarchical triple systems are almost unstable and  disintegrate into a stable binary system and a single star\cite{Stone2019}, and thus stable periodic orbits of non-hierarchical triple systems are rather scarce.   
Here we report one family of 13315 stable periodic orbits of the non-hierarchical triple system with unequal mass. 
Compared with the narrow  mass region (only $10^{-5}$) of the stable figure-eight solution\cite{Galan2002}, our newly-found stable periodic orbits can have fairly large  mass region. 
It is found that many of these newly-found stable periodic orbits have the mass ratios close to those of  the hierarchical  triple systems that have been measured by the astronomical observation.
It implies that these stable periodic orbits of the non-hierarchical triple system with distinctly unequal masses can be quite possibly observed in practice.   Our investigation also suggests that there should exist an infinite number of stable periodic orbits of  non-hierarchical triple systems  with distinctly unequal masses.   Obviously, 
these stable periodic orbits of the non-hierarchical unequal-mass triple system have broad impact for the astrophysical scenario: they could inspire the theoretical and observational study of the non-hierarchical triple system, the formation of triple stars\cite{Reipurth2012}, the gravitational waves pattern\cite{Dmitrasinovic2014} and the gravitational waves observation\cite{Meiron2017} of the non-hierarchical triple system. 
\end{abstract}

Triple systems are common and key objectives in astrophysics\cite{Reipurth2012}.
Although the three-body problem has been investigated for more than three hundred years\cite{Newton1687, Musielak2014}, it is still a challenging and open question for astrophysicist because of its inherent chaotic characteristics\cite{Poincare1890}.
Currently, based on the assumption of ergodicity, Stone and Leigh\cite{Stone2019} gave a statistical solution to the non-hierarchical chaotic three-body system.
It is traditionally believed that bound non-hierarchical triple systems are always unstable and  disintegrate into a stable binary system and a single star\cite{Stone2019}. 
Therefore, periodic orbits of the three-body problem are extremely precious since they are the only way to penetrate the fortress which was previously considered to be inaccessible\cite{Poincare1890}. 
However, only three families of periodic orbits had been found in more than 300 years until \v{S}uvakov and Dmitra\v{s}inovi\'{c}\cite{Suvakov2013} numerically found 13 distinct periodic orbits of the three-body problem with equal mass in 2013. Li and Liao\cite{Li2017-SciChina} further found more than six hundred new families of periodic orbits of the three-body system with equal mass. 
Li et al.\cite{Li2018} gained more than one thousand new families of periodic orbits of the three-body system with two equal-mass bodies. 
Among the about two thousand new families of periodic orbits of the three-body system, dozens of linear stable periodic orbits were found for the non-hierarchical triple system\cite{Dmitravsinovic2018, Li2018}, however, some of them have three equal-mass bodies\cite{Dmitravsinovic2018} and the others have two equal-mass bodies\cite{Li2018}.
The famous figure-eight solution \cite{More1993, Chenciner2000} of the equal-mass triple system is non-hierarchical and linear stable \cite{Simo2002}. Unfortunately,  the stable mass region of the figure-eight solution is very narrow (only $10^{-5}$) \cite{Galan2002}. That is to say the figure-eight solution is stable only when three bodies have almost equal mass, so the probability of observing this periodic orbit is extremely low in practice. So far,
{\em non-hierarchical} periodic triple stars have {\em not} been found  in the astronomical observation yet. 

In this letter, we focus on periodic orbits of non-hierarchical triple system with {\em unequal} masses.
The motion of the Newtonian planar three-body problem is described by the differential equations
\begin{equation}
\ddot{\bm{r}}_{i}=\sum_{j=1,j\neq i}^{3} \frac{G m_{j}(\bm{r}_{j}-\bm{r}_{i})}{| \bm{r}_i-\bm{r}_j |^{3}},\label{ODE}
\end{equation}
where $m_i$ and ${\bm r}_i$ are mass and position of the $i$th body $(i=1,2,3)$,  $G$ is the Newtonian gravity constant, respectively.  Without loss of generality, we set the gravitational constant $G=1$ by properly choosing  a  characteristic mass $M$, a characteristic spatial length $R$ and a characteristic time $T^*$. 

Montgomery\cite{Montgomery2007} proofed that all three-body orbits of zero angular momentum have syzygies (i.e., collinear instant of three bodies) except for the  Lagrange's solution. Thus, it is reasonable to consider initial conditions with the collinear configuration\cite{Hadjidemetriou1975, Henon1976, Jankovic2020}.
In this letter, we investigate the unequal-mass triple system with the initial positions
$\bm{r}_1(0)=(x_1,0)$, $\bm{r}_2(0)=(x_2,0)$, $\bm{r}_3(0)=(x_3,0)$ 
and the initial velocities  $\dot{\bm{r}}_1(0)=(0,v_1)$, $\dot{\bm{r}}_2(0)=(0,v_2)$, $\dot{\bm{r}}_3(0)=(0, v_3)$,  which are perpendicular to the straight line formed by three bodies. 

The first step to achieve our goal is to find periodic orbits of the equal-mass triple system with the collinear initial condition configuration mentioned above.
We numerically search for periodic orbits of the three-body problem with equal mass and zero angular momentum by means of the grid search method, the Newton-Raphson method\cite{Farantos1995, Lara2002} and the numerical strategy, namely the clean numerical simulation (CNS)\cite{Liao2009, Liao2014,Liao2014-SciChina, Hu2020} (see Methods). 
We find that one equal-mass periodic orbit has good stability.
The initial condition of this periodic orbit is $\bm{r}_1(0)=(x_1, 0)$, $\bm{r}_2(0)=(1, 0)$, $\bm{r}_3(0)=(0, 0)$, $\dot{\bm{r}}_1(0)=(0, v_1)$, $\dot{\bm{r}}_2(0)=(0, v_2)$,  $\dot{\bm{r}}_3(0)=(0,-(m_1v_1+m_2v_2)/m_3)$, where $x_1 = -0.372008640907423$, $v_1 =  1.21800411067968$, $v_2=0.4531080538336022$ and the period $T = 7.53971451331772$ and $m_1=m_2=m_3=1$. 
Note that it holds $m_1 v_1 x_1 + m_2 v_2 x_2 + m_3 v_3 x_3 = 0$.  
Using the homotopy classification method\cite{Montgomery1998, Suvakov2013}, the free group element of this periodic orbit is $bABabaBAba$. This periodic orbit has the same free group element with the moth-I orbit\cite{Suvakov2013}, but their periodic orbits are different.
Note that, for the astrophysical three-body system, the masses of the bodies are rarely equal. Thus, using this as a starting point, we investigate periodic orbits of the unequal-mass triple  system by means of the numerical continuation method\cite{Allgower2003} (see Methods).

\begin{figure}
\includegraphics[scale=0.25]{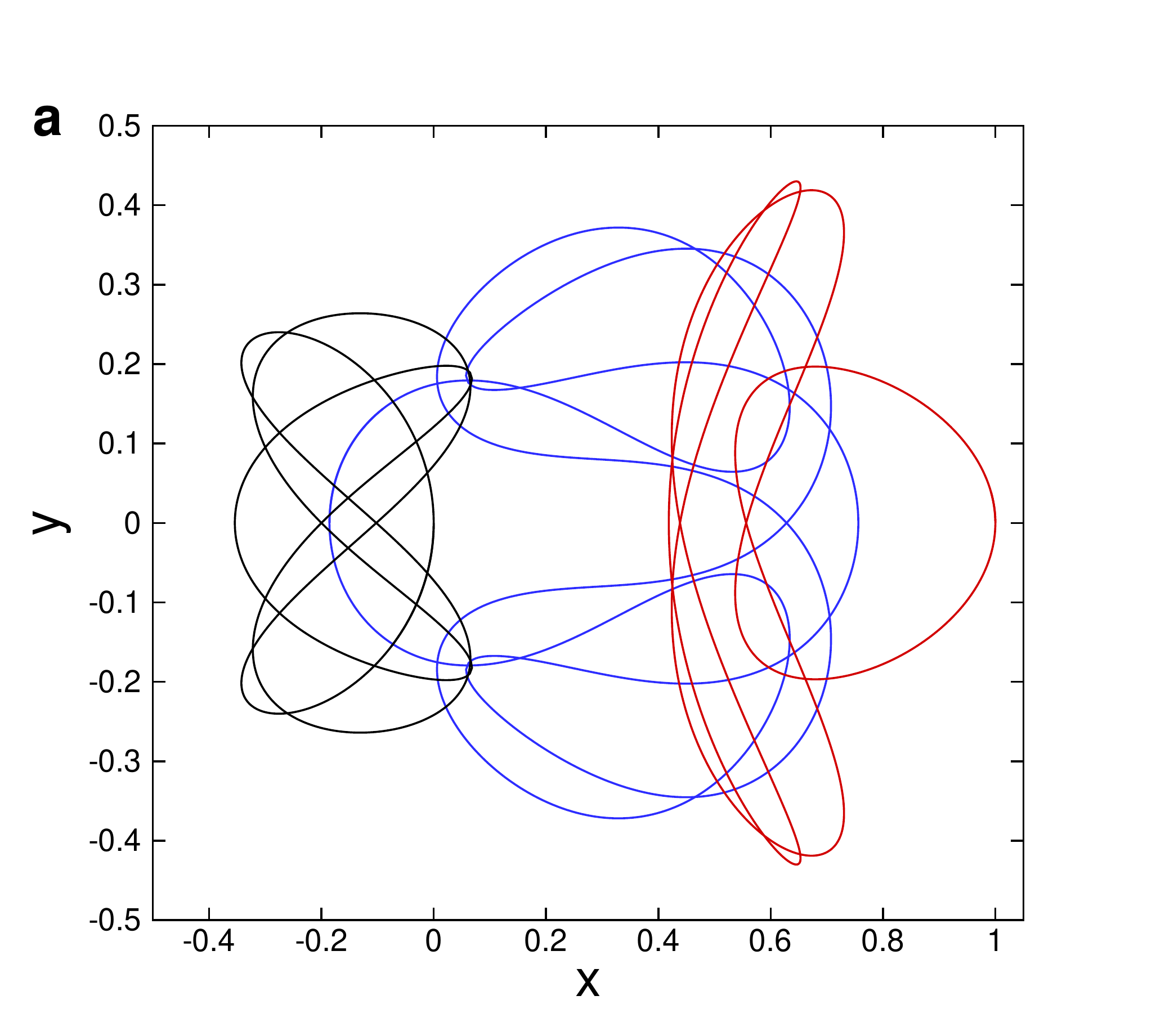}
\includegraphics[scale=0.25]{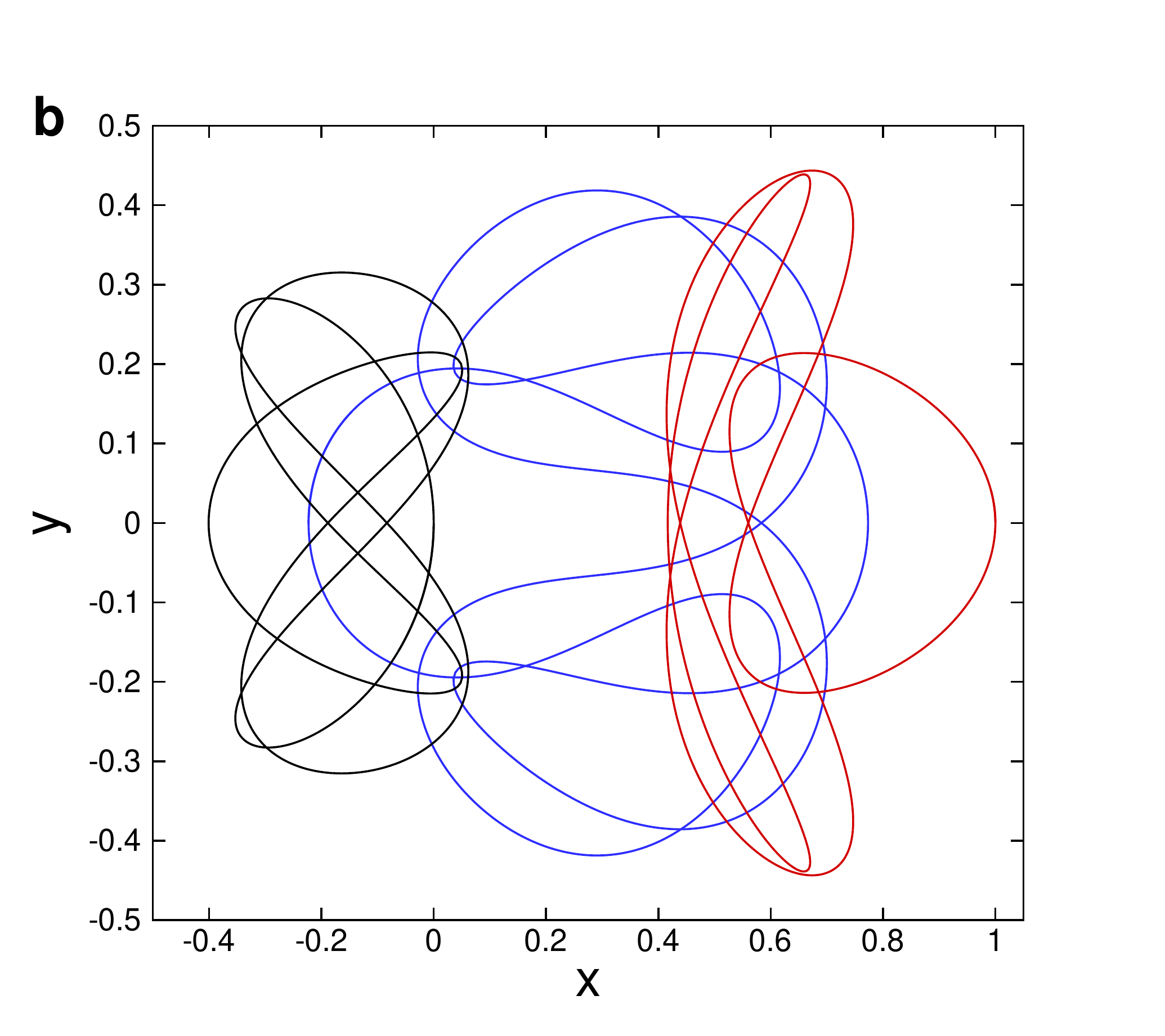}
\includegraphics[scale=0.25]{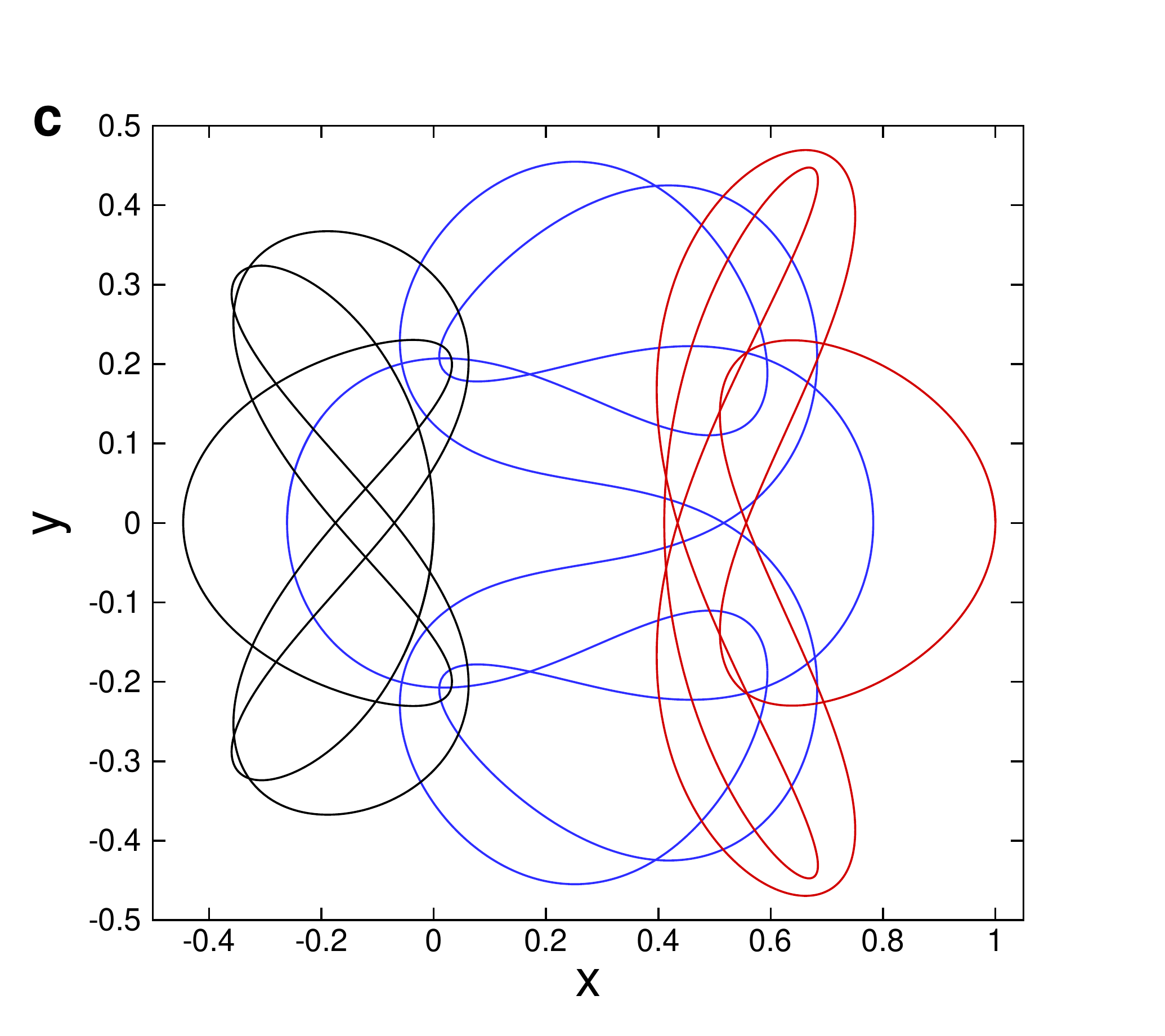}
\caption{Three newly-found stable periodic orbits of the non-hierarchical triple systems with different masses and period: (a) $m_1=0.87$, $m_2=0.8$, $m_3=1$ and $T = 5.9889127121$; (b) $m_1 = 0.9$, $m_2 =0.85$, $m_3=1$ and $T = 6.3508660391$; (c) $m_1=0.93$, $m_2=0.89$,  $m_3=1$ and $T=6.6805531109$.  Body-1: blue line; Body-2: red line; Body-3: black line.  }
\label{stable-orbit}
\end{figure}

Starting from the periodic orbit of the equal-mass triple system mentioned above, we obtain $135445$ periodic orbits in the region of $m_1 \in [0.8, 1.1]$ and $m_2 \in [0.7, 1.2]$ with a fixed mass $m_3=1$ by means of the continuation method (see Methods).
The periodic orbits are outputted with the mass interval $\delta m = 0.001$.
The detailed initial conditions and periods are listed in the supplementary data.
Three examples of these periodic orbits are shown in Figure~\ref{stable-orbit}. Their initial conditions and periods of the three periodic orbits are listed in Table~\ref{stable-ini}.

\begin{table}
\tabcolsep 0pt \caption{Initial conditions and periods $T$ of three stable periodic orbits for the non-hierarchical three-body system in the case of  $\bm{r}_1(0)=(x_1,0)$, $\bm{r}_2(0)=(1,0)$, $\bm{r}_3(0)=(0,0)$, $\dot{\bm{r}}_1(0)=(0,v_1)$, $\dot{\bm{r}}_2(0)=(0,v_2)$,  $\dot{\bm{r}}_3(0)=(0,-(m_1v_1+m_2v_2)/m_3)$ when $G=1$.} \label{stable-ini} \vspace*{-12pt}
\begin{center}
\def\temptablewidth{1\textwidth}
{\rule{\temptablewidth}{1pt}}
\begin{tabular*}{\temptablewidth}{@{\extracolsep{\fill}}lllcccc}
$m_1$ & $m_2$ & $m_3$ & $x_1$ & $v_1$  & $v_2$  & $T$ \\
\hline
0.87	&	0.8	&	1	&	-0.1855174644	&	2.0221546880	&	0.3968976468	&	5.9889127121	\\
0.9	&	0.85	&	1	&	-0.2227468469	&	1.7812769516	&	0.4150035570	&	6.3508660391	\\
0.93	&	0.89	&	1	&	-0.2610366744	&	1.5883335319	&	0.4304477015	&	6.6805531109	\\
\end{tabular*}
{\rule{\temptablewidth}{1pt}}
\end{center}
\end{table}

\begin{figure}
\includegraphics[scale=0.4]{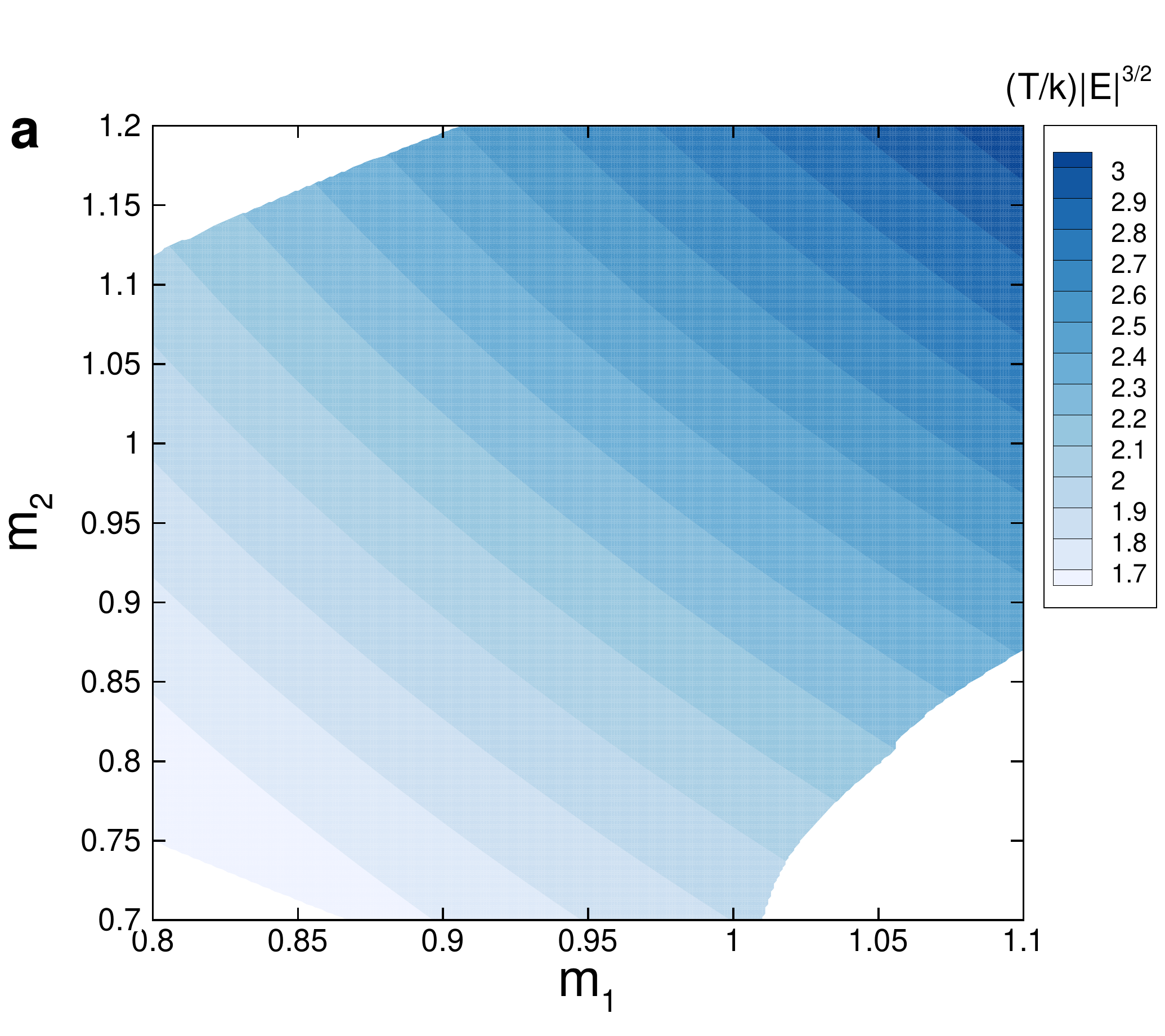}
\includegraphics[scale=0.4]{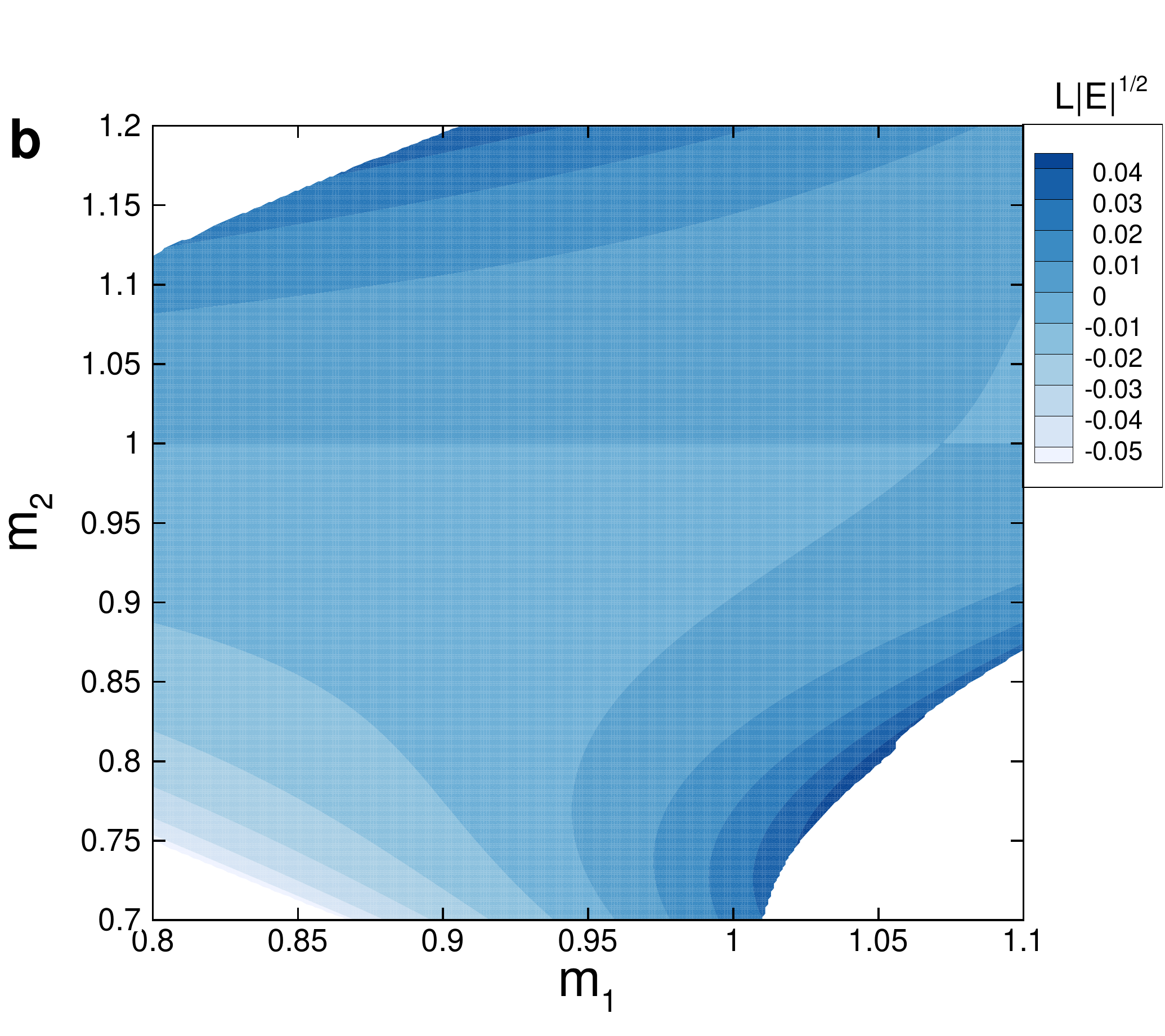}
\caption{The contour map of the scale-invariant average period and scale-invariant angular momentum of newly-found periodic orbits: (a) The contour map of the average scale-invariant period $\bar{T}^* = (T/k) |E|^{3/2}$ in the $m_1$-$m_2$ plane, where $E$, $T$, $k$ is total energy, period and the number of free group words of periodic orbits, respectively; (b) The contour map of the scale-invariant angular momentum $L|E|^{1/2}$  in the $m_1$-$m_2$ plane, where $L$ is angular momentum.}
\label{kepler}
\end{figure}

Due to the homogeneity of the potential field for the three-body problem, there is a scaling law: $\bm{r'}=\alpha\bm{r}$, $\bm{v'}=\bm{v}/\sqrt{\alpha}$, $t'=\alpha^{3/2}t$ and energy $E'=E/\alpha$ and angular momentum $L'=\sqrt{\alpha}L$. 
The scale-invariant average  period $\bar{T}^* = (T/k) |E|^{3/2}$ is approximately equal to a constant for periodic orbits of the three-body problem with equal mass\cite{Dmitrasinovic2015, Li2017-SciChina}, where $k$ is the number of free group words of periodic orbits. 
For the family of periodic orbits $bABabaBAba$, we always have the number of the free group words $k=10$.
For the newly-found periodic unequal-mass orbits, Figure~\ref{kepler}a shows that the scale-invariant average period $\bar{T}^* = (T/k) |E|^{3/2}$ depends on the mass of bodies. 
The multiple linear regression for these periodic orbits is $(T/k)|E|^{3/2} = 2.455 m_1+ 1.655 m_2 - 1.688$. The standard error of this multiple linear regression is $0.021$. It indicates that the  scale-invariant average period $\bar{T}^* = (T/k) |E|^{3/2}$ is approximately linear to $m_1$ and $m_2$ for this family of periodic orbits. 
Jankovi\'{c} and Dmitra\v{s}inovi\'{c}\cite{Jankovic2016} found that the scale-invariant angular momentum is a function of topologically rescaled period for the Broucke-Hadjidemetriou-H\'{e}non family of periodic triple orbits with equal mass.
For our newly-found family of periodic orbits, it is demonstrated that the scale-invariant angular momentum $L|E|^{1/2}$ varies among different masses $m_1$ and $m_2$ as shown in Figure~\ref{kepler}b. It implies that the scale-invariant angular momentum also depends on the mass of bodies for this family of periodic orbits of the unequal-mass triple system.
Note that some regions of the Figure~\ref{kepler} is blank. It suggests that no periodic orbits can be found there because the orbits of the three-body system might have collision in that mass region. 
 
Stability is an important property for periodic orbits because only stable triple system can probably be observed.
The stability of periodic orbits of the three-body system can be investigated according to the characteristic multipliers of the monondromy matrix\cite{Simo2002}. Due to the fixed center of mass, the dimension of the planar three-body problem can be reduced to eight.
We employ a theorem proofed by Kepela and Sim\'{o}\cite{kapela2007} to determine the linear stability of periodic orbits of three-body problem through the monondromy matrix.
With the monodromy matrix, we can gain the equation as follows:
\begin{equation}
T^2-(\alpha-4)T+\beta-4\alpha+8=0,
\end{equation} \label{stability}
where $\alpha = trace(A) =\sum_{i=1}^8a_{ii}$, $\beta=\sum_{1\leq i<j \leq 8}(a_{ii}a_{jj}-a_{ij}a_{ji})$, $a_{ij}$ is the elements of the monondromy matrix $A$.

$\textbf{Therorem}$\cite{kapela2007}. Let $T_1$ and $T_2$ be solutions of the equation (2). If $\Delta = (\alpha-4)^2-4(\beta-4\alpha+8)>0$, $|T_1|<2$ and $|T_2|<2$, then all eigenvalues of the monodromy matrix $A$ are on the unit circle.

\begin{figure}
\centering \includegraphics[scale=0.45]{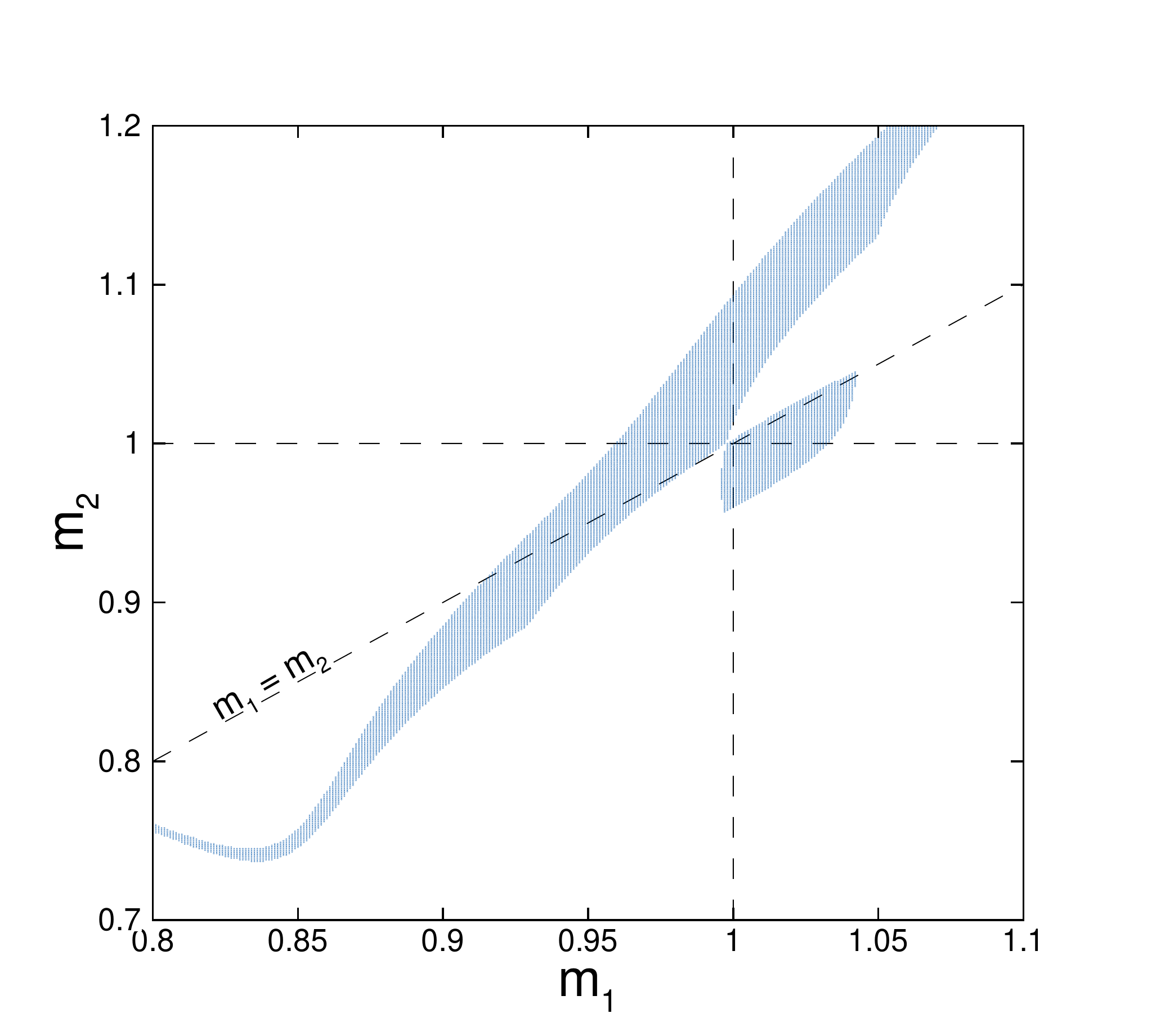}
\caption{The stability region of periodic orbits in the $m_1$-$m_2$ plane. Shadowing domain: stable periodic orbits. }\label{stability-region}
\end{figure}

Using this theorem, we find that $13315$ periodic orbits are linear stable among the $135445$ newly-found periodic orbits.
Three examples of the stable periodic orbits are shown in Table~\ref{stable-ini} and Figure~\ref{stable-orbit}. 
The domain of the masses $(m_1, m_2)$ of stable periodic orbits is shown in Figure~\ref{stability-region}. 
The mass region becomes narrow when the masses $m_1$ and $m_2$ decrease.
Notice that the mass region of the stable figure-eight solution\cite{More1993, Chenciner2000} is very narrow (only $10^{-5}$)\cite{Galan2002}.
So, the mass region of the newly-found stable non-hierarchical periodic orbits is fairly large and their masses have apparent differences. 
For instance, for the stable {\em non-hierarchical} periodic orbit $m_1=0.87$, $m_2=0.8$ and $m_3=1$, we have its mass ratio $m_2/m_1\approx 0.92$ and $m_2/m_3 = 0.8$.
A recent observed {\em hierarchical} triple system\cite{Dimitrov2017} has masses $1.21$, $1.14$ and $1.4$ $M_{\odot}$, corresponding to mass ratios $0.94$ and $0.81$.
It should be emphasized that the mass ratios of our newly-found stable {\em non-hierarchical} periodic orbits are close to the mass ratios of the  hierarchical  triple system which has been measured by the astronomical observation. This implies that our newly-found stable {\em non-hierarchical} periodic orbits are likely to be observed in astronomy.

Since the dimensionless quantities are used in above numerical results,
the variables can be rescaled to applications of stellar dynamics\cite{Szebehely1967} through  $GMT^{*^2}/R^3=1$, where $M$, $T^*$ and $R$ is the characteristic mass, time and length and $G$ is the Newtonian gravitational constant. 
If we choose $M=M_{\odot}$ and $R=10$ AU, 
then $T^*=\sqrt{\frac{R^3}{GM}}\approx 5 $ years.
For instance, with these units of quantities, the stable {\em non-hierarchical} periodic orbit with $m_1=0.87 M_{\odot}$, $m_2= 0.8M_{\odot}$ and $m_3= M_{\odot}$ has period for about $30$ years. Note that the hierarchical triple system HD 188753 has period for 25 years and semi-major axis for 11.8 AU\cite{Marcadon2018}. Thus, our newly-found stable {\em non-hierarchical} triple systems have similar size and period with the observed {\em hierarchical} triple system. 

There may be two reasons why the non-hierarchical periodic triple stars has not been found in the astronomical observation yet.
On one hand, the accurate positions and motions of the non-hierarchical systems were not easy to determine because they are complicated and far away from the earth.
On the other hand, there were few periodic non-hierarchical unequal-mass triple systems found in theoretical and numerical study before.
Fortunately, Gaia mission\cite{Prusti2016} has produced high-precision measurements of positions and motions of nearly 1.7 billion stars which provide  
resource to study non-hierarchical periodic triple systems. This implies that our newly-found stable {\em non-hierarchical} periodic orbits are likely to be observed in near future.

In this letter, we present one family of $135445$ periodic orbits for non-hierarchical triple system with unequal masses. 
Surprisely, among these $135445$ periodic orbits of this family, $13315$ periodic orbits are linear stable in a large mass region. 
Most of them have fairly different masses, which implies that our newly-found stable periodic orbits are likely to be observed in practice.
Note that we only consider here one family of the periodic orbits with the free group element $bABabaBAba$, but found $13315$ stable ones among the $135445$ periodic orbits.  Note also that hundreds of families of periodic equal-mass three-body orbits were found\cite{Li2017-SciChina}: similarly, each of them as a starting point might lead to thousands of stable periodic orbits of the  non-hierarchical triple system with unequal-mass.  Therefore, in theory,  there  should exist an infinite  number   of   stable  periodic orbits of non-hierarchical  triple systems with distinctly unequal mass.     
Our newly-found stable periodic orbits of the non-hierarchical unequal-mass triple system have broad impact for the astrophysical scenario: they could inspire the theoretical and observational study of the non-hierarchical triple system, the formation of triple stars\cite{Reipurth2012}, the gravitational waves pattern\cite{Dmitrasinovic2014} and the gravitational waves observation\cite{Meiron2017} of the non-hierarchical triple system. 

\begin{methods}
\subsection{Clean Numerical Simulation.} The clean numerical simulation (CNS) \cite{Liao2009, Liao2014-SciChina, Liao2014, Hu2020} is a numerical strategy to gain reliable numerical simulation of chaotic dynamical systems, such as the three-body system. The CNS is based on an arbitrary Taylor series method \cite{Corliss1982, Chang1994, Barrio2005} and multiple-precision arithmetic\cite{Oyanarte1990}, plus a convergence verification by means of an additional computation with smaller numerical noise. 
Li and Liao\cite{Li2017-SciChina, Li2019} found that many periodic orbits of three-body problem might be lost by
using conventional numerical algorithms in double precision.
Thus, here we apply the CNS to integrate the differential equations of the three-body system. 

\subsection{Numerical searching method.}
At the beginning, we numerically search for periodic orbits of the three-body problem with equal masses $m_1=m_2=m_3=1$ and zero angular momentum. Due to the homogeneity of the potential field for the three-body problem, the initial condition $x_2$ can be fixed to unit. Then we choose the velocity $v_2=-x_1v_1$ due to  zero angular momentum. Without loss of generality, we assume total momentum $m_1\dot{\bm{r_1}}+m_2\dot{\bm{r_2}}+m_3\dot{\bm{r_3}}=0$. Therefore, the initial positions can be specified as
$\bm{r}_1(0)=(x_1,0)$, $\bm{r}_2(0)=(1,0)$,   $\bm{r}_3(0)=(0,0)$
and the initial velocities can be specified as
$\dot{\bm{r}}_1(0)=(0,v_1)$,  $\dot{\bm{r}}_2(0)=(0,-x_1v_1)$,  $\dot{\bm{r}}_3(0)=(0, -v_1+x_1v_1)$.

With the initial configuration, the orbits of the three-body problem are determined by two parameters $x_1$ and $v_1$. According to the numerical searching method of the three-body problem\cite{Suvakov2013, Li2017-SciChina}, the first step is to gain approximated initial values of periodic orbits in a two dimensional space (i.e., the $x_1$-$v_1$ plane). We investigate a region of this plane: $x_1\in (-1,0)$ and $v_1\in (0,10)$.
We employ $4000 \times 40000$ uniform grid points as initial conditions in this region.  With these initial conditions, the differential equations (\ref{ODE})  are numerically solved by an eight-oder Runge Kutta ODE solver dop853 developed by Hairer et al.\cite{Hairer1993}.
For each initial condition, the return proximity function $d(\bm{y}(0), T_0) = \min\limits_{t \leq T_0}||\bm{y}(t)-\bm{y}(0)||$ is calculated up to integration time $T_0=200$.
We choose the initial conditions and periods $T$  as possible candidates of periodic orbits when the return proximity function $d(\bm{y}(0), T_0)<0.1$.

The next step is to improve the precision of the approximate initial conditions of the periodic orbits using the Newton-Raphson method \cite{Farantos1995, Lara2002}  and the clean numerical simulation (CNS) by means of  correcting the parameters $x_1$, $v_1$ and period $T$. The precision of the initial conditions of the periodic orbits is improved continually until the level of the return proximity function is less than
$10^{-12}$.

\subsection{Continuation method.}

The numerical continuation method\cite{Allgower2003} is used to gain periodic solutions of a nonlinear dynamical system  with a natural parameter
\begin{equation}
\dot{\bm{u}} = G(\bm{u},\lambda).
\end{equation} 
Using a known periodic orbit $\bm{u}_0$ at $\lambda_0$ as initial guess, we can obtain a new periodic orbit $\bm{u}'$ at $\lambda+\Delta \lambda$ by means of the Newton-Raphson method\cite{Farantos1995, Lara2002} and the clean numerical simulation (CNS)\cite{Liao2009, Liao2014-SciChina, Liao2014, Hu2020} when $\Delta \lambda$ is sufficient small  to guarantee the convergence of iteration.

Because of homogeneity of the potential field of the three-body problem, we can fix the initial distance of two bodies to unit. Without loss of generality, we consider the case of zero  momentum (i.e., $m_1\dot{\bm{r_1}}+m_2\dot{\bm{r_2}}+m_3\dot{\bm{r_3}}=0$). The periodic orbits are determined by $x_1$, $v_1$, $v_2$ and $T$ with masses $m_1$, $m_2$ and $m_3$. Therefore, the initial positions of three bodies can be described by
\begin{equation}
\bm{r}_1(0)=(x_1,0), \;\; \bm{r}_2(0)=(1,0), \;\;  \bm{r}_3(0)=(0,0), 
\end{equation}
and the initial velocities can be described by
\begin{equation}   
 \dot{\bm{r}}_1(0)=(0,v_1), \;\; \dot{\bm{r}}_2(0)=(0,v_2), \;\; \dot{\bm{r}}_3(0)=(0, -\frac{m_1v_1+m_2v_2}{m_3}).
\end{equation}

With the fixed masses $m_2=m_3=1$, periodic orbits can be obtained by means of the numerical continuation method for different mass $m_1$.
Using a periodic orbit with equal mass as a starting point, we apply the Newton-Raphson method and the clean numerical simulation (CNS) to gain a new periodic orbit at $m_1+\Delta m$  by continually modifying the parameters $x_1$, $v_1$, $v_2$ and $T$, where $\Delta m$ is small enough to guarantee the convergence of iteration.  In this way, we can gain periodic orbits with different mass $m_1\neq 1$ and $m_2=m_3=1$.

Similarly, using the above periodic orbits with $m_1\neq 1$ and $m_2=m_3=1$ as starting points, we further employ the Newton-Raphson method and the clean numerical simulation (CNS) to gain periodic orbits at $m_2+\Delta m$  by continuously correcting the parameters $x_1$, $v_1$, $v_2$ and $T$, where $\Delta m$ is small enough to guarantee the convergence of iteration.  Consequently, we gain periodic orbits of the triple system with unequal masses $m_1 \neq m_2 \neq m_3$. 

Note that the periodic orbits might have nonzero angular momentum since we don't restrict the angular momentum.

\end{methods}


\bibliography{ref}


\begin{addendum}
 \item This work was carried out on TH-1A at National Supercomputer Center in Tianjin and TH-2 at National Supercomputer Center in Guangzhou, China.  It is partly supported by National Natural Science Foundation of China (Approval No. 11702099 and 91752104) and the International Program of Guangdong Provincial Outstanding Young Researcher.
 \item[Author contributions] X.M.L. calculated the periodic orbits and generated the first draft. X.C.L. analysed the data, discussed the results and modified the letter. S.J.L. discussed the results and revised the letter. All authors contributed to the discussion and revision of the final manuscript.
 \item[Competing Interests] The authors declare that they have no
competing financial interests.
 \item[Correspondence] Correspondence and requests for materials
should be addressed to S.J.L. (sjliao@sjtu.edu.cn) or X.C.L. (xiaochenli@scut.edu.cn). 
\end{addendum}


\end{document}